\documentclass[prb,twocolumn,superscriptaddress, floatfix]{revtex4-2}

\usepackage{graphicx,color}
\usepackage{tikz}
\usepackage{amssymb}
\usepackage{amsmath,amsfonts,amsthm} 
\usepackage{natbib}
\usepackage[colorlinks=true,linkcolor=blue,citecolor=blue,urlcolor=blue]{hyperref}
\usepackage{ulem}
\usepackage{physics}
\usepackage{wasysym}

\newcommand{\hexagona}{\tikz[baseline=0.7ex]{
\fill (0,1.3ex) coordinate (A);
\fill (0.75ex,2.6ex) coordinate (B);
\fill (2.25ex,2.6ex) coordinate (C);
\fill (3ex,1.3ex) coordinate (D);
\fill (2.25ex,0ex) coordinate (E);
\fill (0.75ex,0ex) coordinate (F);
\draw (A)--(B)--(C)--(D)--(E)--(F)--(A);
\fill (0.3ex,1.3ex) coordinate (G);
\fill (0.9ex,2.34ex) coordinate (H);
\fill (2.1ex,2.34ex) coordinate (I);
\fill (2.7ex,1.3ex) coordinate (J);
\fill (2.1ex,0.26ex) coordinate (K);
\fill (0.9ex,0.26ex) coordinate (L);
\draw (G)--(H);
\draw (I)--(J);
\draw (K)--(L);
}}

\newcommand{\hexagonb}{\tikz[baseline=0.7ex]{
\fill (0,1.3ex) coordinate (A);
\fill (0.75ex,2.6ex) coordinate (B);
\fill (2.25ex,2.6ex) coordinate (C);
\fill (3ex,1.3ex) coordinate (D);
\fill (2.25ex,0ex) coordinate (E);
\fill (0.75ex,0ex) coordinate (F);
\draw (A)--(B)--(C)--(D)--(E)--(F)--(A);
\fill (0.3ex,1.3ex) coordinate (G);
\fill (0.9ex,2.34ex) coordinate (H);
\fill (2.1ex,2.34ex) coordinate (I);
\fill (2.7ex,1.3ex) coordinate (J);
\fill (2.1ex,0.26ex) coordinate (K);
\fill (0.9ex,0.26ex) coordinate (L);
\draw (H)--(I);
\draw (J)--(K);
\draw (L)--(G);
}}

\begin{document}

\title{Neutron scattering from fragmented frustrated magnets}

\author{F. Museur}
\affiliation{Institut N\'eel, CNRS \& Univ. Grenoble Alpes, 38042 Grenoble, France}
\affiliation{ENS de Lyon, CNRS, Laboratoire de Physique, F-69342 Lyon, France}
\author{E. Lhotel}
\affiliation{Institut N\'eel, CNRS \& Univ. Grenoble Alpes, 38042 Grenoble, France}
\author{P. C. W. Holdsworth}
\affiliation{ENS de Lyon, CNRS, Laboratoire de Physique, F-69342 Lyon, France}

\begin{abstract}
    The fragmentation description is used to analyse calculated neutron scattering intensities from  kagom\'e ice and spin ice systems. The longitudinal, transverse and harmonic fragments produce independent contributions to the neutron scattering intensity. This framework is used to analyse the ordering due to quantum fluctuations in the topologically constrained phase of kagom\'e ice and the monopole crystal phase of spin ice. Here, quantum fluctuations are restricted to the transverse fragment and they drive the system into a double-$q$ structure in which longitudinal and transverse fragments have a different ordering wave vector. The intensity reduction of the Bragg peaks for the transverse fragments, compared with known classical limits can be used as a diagnostic tool for quantum fluctuations. Published quantum Monte Carlo data for spin ice in a $[111]$ field are consistent with the proposed protocol.
\end{abstract}

\maketitle

\section{Introduction}
Emergent gauge field descriptions \cite{Isakov2004} have revolutionised our vision of frustrated magnetism, leading us far from our expectations for microscopic systems. The monopole picture \cite{Castelnovo2008,Ryzhkin2005} of spin ice \cite{harris1997geometrical,bramwell2001spin},    the $U(1)$ quantum spin liquid phase \cite{Hermele2004,Benton2012,Balents2010,Gingras2014}, or more recently the possible emergence of gauge fields of higher rank \cite{Yan2020} are particularly remarkable examples. In spin ice this emergent description is a good approximation even at the microscopic level \cite{Castelnovo2021modelling,Kaiser2018,Castelnovo2011,Fennell2009} so that the magnetic moments represent elements of a lattice field which, at low temperature and in zero external field is the curl of an emergent gauge field, $\vec {A}$ \cite{Isakov2004,Brooks2014}. This so called ``transverse'' field leads to dipolar spin correlations in the low temperature ``Coulomb'' phase \cite{Henley2010}. The excitations of magnetic monopoles out of this monopole vacuum requires the syphoning off of a part of this magnetic flux reservoir to create an orthogonal, or ``longitudinal''  field, the gradient of a scalar potential $\Psi$. Application of an external magnetic field or the presence of surface charges requires a further separation giving the required harmonic field contribution \cite{Bramwell2017harmonic,Bhatia2013}, $\vec {h}$. This fragmentation of the magnetic resources \cite{Brooks2014,Lhotel2020fragmentation} corresponds to a Helmholtz-Hodge decomposition \cite{Bhatia2013} of the emergent vector field
\begin{equation}
    \vec {M} = \grad{\Psi} + \curl{\vec {A}} + \vec {h}.\label{Helmholtz}
\end{equation}

In this paper we explore the consequences of fragmentation for neutron scattering on spin ice like systems, concentrating on those in which quantum fluctuations are in competition with, or responsible for the development of long rang magnetic order. In these systems, because of the separation in energy scales associated with the transverse and longitudinal fragments, quantum fluctuations are largely restricted to the transverse fragment. As a result, the fragmentation picture is extremely useful for the analysis. A characteristic of the ordered phases discussed is that the longitudinal and transverse fragments order with different wave vector giving examples of ``double-$q$'' structures \cite{Zhitomirsky2022}. As our main example we concentrate on order driven from the ``KII'', topological liquid phase of kagome ice \cite{Moller2009,Chern2011,Wang2020}. We show that, while the longitudinal fragment responsible for the charge orders in a $q=0$ structure, the transverse fragment orders at a finite wave vector characteristic of the ``dimer star phase'' defined in detail below.
The KII phase is generated in models of two-dimensional kagom\'e ice that include long range magnetostatic interactions and in the kagom\'e spin planes lying perpendicular to an external field placed along the $[111]$ direction in a spin ice material.
Ordering out of the KII phase can be driven either classically by potential energy  \cite{Moller2009,Chern2011}, or by quantum fluctuations  in quantum spin models \cite{Wu2019,Bojesen2017}. We make predictions for neutron scattering intensities from classical and quantum ordered states and compare with published numerical data from quantum Monte Carlo simulations on quantum spin ice in a $[111]$ field \cite{Bojesen2017}. We also comment on the analogous three dimensional system, the monopole crystal phase of quantum spin ice \cite{Jaubert2015} in which a dense, ordered monopole structure would cohabit with a quantum spin liquid superposition of the transverse fragments.

The problems considered map onto dimer problems via their emergent field description \cite{Huse2003}. In these phases, which  show magnetic charge order, the transverse fragment maps exactly onto one of the $Z_2$ sectors of these emergent fields. The magnetic charge ordering explicitly breaks this $Z_2$ symmetry leaving a unique opportunity to observe dimer physics, quantum or classical, with a dipolar probe. That is, using neutron scattering within the dipole approximation one can visualise the correlations emerging from the fictive quadrupolar objects.

The rest of the paper is organised as follows: in the next two sections we review the fragmentation picture for spin-ice and its two-dimensional equivalent. In section IV we show that neutron scattering data conveniently splits into identifiable contributions from the transverse, longitudinal and harmonic fragments. We illustrate this discussion using the low temperature, ``$\sqrt{3}\times\sqrt{3}$'' phase of classical dipolar kagom\'e ice. In section V we show how quantum fluctuations quantitatively change the predicted neutron scattering patterns as the classical phase changes to a quantum resonating phase and illustrate how this can be used as a diagnostic tool for detecting quantum fluctuations. In section VI we relate our discussion to published quantum Monte Carlo data \cite{Bojesen2017}. Section VII deals with the monopole crystal in spin ice and the paper concludes with a general discussion. Throughout the paper we refer to the kagom\'e plateau region of spin ice with an applied $[111]$ field as the kagom\'e plateau and to the two-dimensional problem of a single kagom\'e plane of triangles as kagom\'e ice.

\section{A review of Fragmentation}
\begin{figure}[h]
    \centering
    \includegraphics[width=0.8\linewidth]{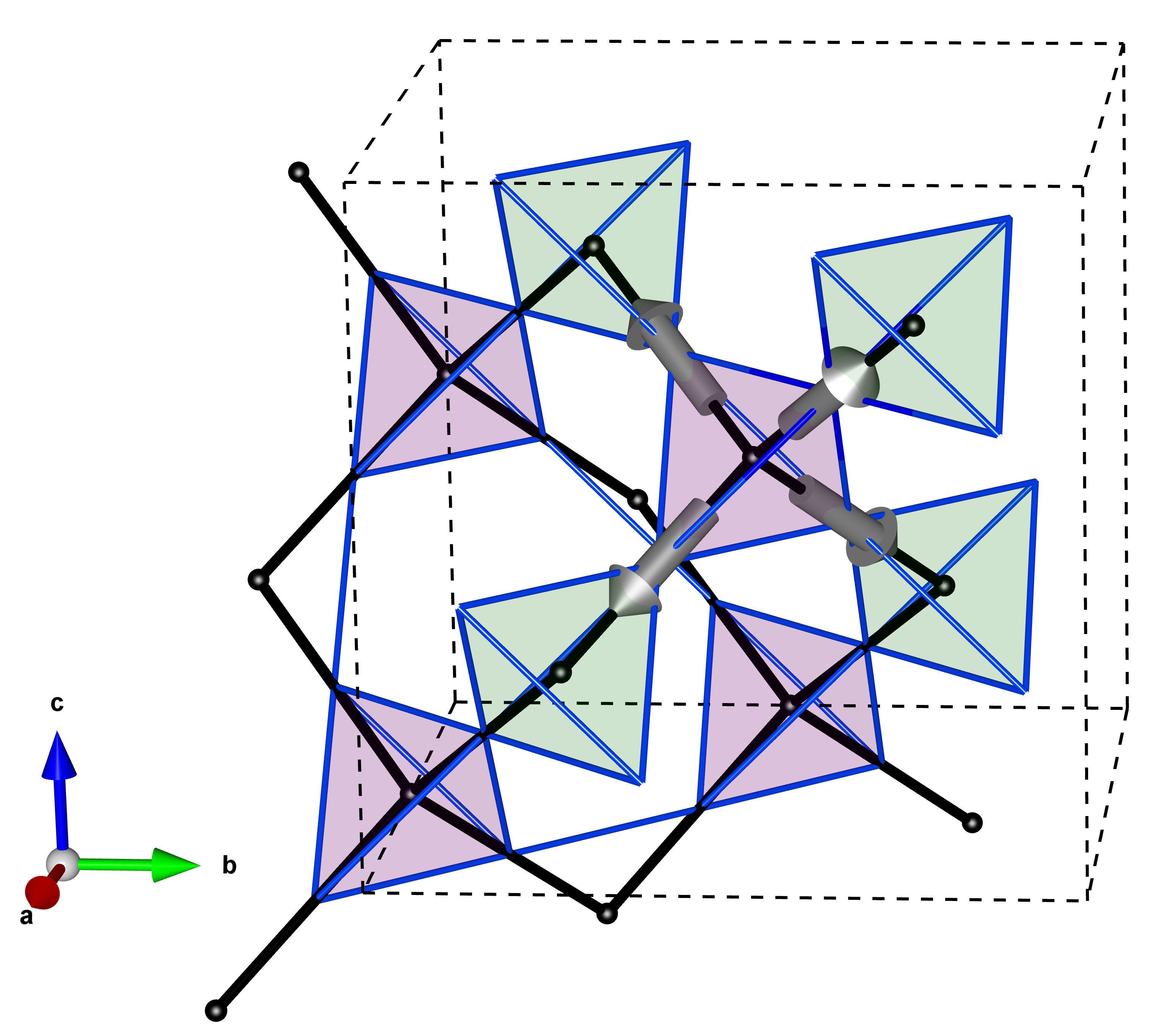}
    \caption{Pyrochlore lattice (blue) and its dual diamond lattice (black). Tetrahedra of type $A$ ($B$) are shown in shaded purple (green) respectively. The cubic unit cell is delimited by the dashed lines and contains sixteen sites. The grey arrows show the four $\vec d_i$ vectors defined in the text. }
    \label{pyrochlore-lattice}
\end{figure}

Spin ice forms a pyrochlore lattice of corner sharing tetrahedra, a four-sublattice face centred cubic structure. The convention in discussing this system is to use the overlying cube of side $a_c$ containing  sixteen sites,  as shown in Fig. (\ref{pyrochlore-lattice}). A laboratory frame $[\hat{x},\hat{y},\hat{z}]$ is then defined with respect to the basis vectors of the cube.
The  spins take discrete orientations, pointing towards or away from the centres of the tetrahedra, along one of the four body diagonals of the cube: $\vec S_i=\pm \vec d_{i}$
\begin{align}
\label{vertexorder}
    \vec d_{1} & = \frac{1}{\sqrt{3}}[-1,-1,1], \quad   &\vec d_2&=\frac{1}{\sqrt{3}}[1,-1,-1] \nonumber \\
    \vec d_3   & = \frac{1}{\sqrt{3}}[-1,1,-1],  \quad &\vec d_4&=\frac{1}{\sqrt{3}}[1,1,1]. 
\end{align}
At the microscopic level, the monopole picture corresponds to replacing the spins by needles carrying magnetic flux and therefore dumbbells \cite{Moller2006,Castelnovo2008} of magnetic charges into the centres of the tetrahedra. The tetrahedra form a diamond lattice of magnetic charge vertices, with a spin on each bond.
The spins and diamond lattice sites are labeled $i,j$ and $I,J$ respectively.
The needles carry flux units of $m/a$, where $m$ is the magnetic moment associated with the spin and $a=\frac{\sqrt{3}}{4} a_c$ the diamond lattice constant. They can thus be considered as elements of a lattice field lying along the bonds of the diamond lattice:
$M_{IJ}=(\vec S_i.\vec d_i) \frac{m}{a}\eta_I$.
For the bipartite diamond lattice, $\eta_I=1$ for a tetrahedron of type $A$ in which the out pointing spin $\vec S_1$ falls along $\vec d_{1}$ and  $\eta_I=-1$ for type $B$ which is the inverse. This convention ensures that $M_{IJ}=-M_{JI}$. These scalar elements can be converted into vector field elements by multiplying once again by the unit vector $\vec d_i$ lying on the bond $IJ$, $\vec M_{IJ}=M_{IJ} \vec d_i=-\vec M_{JI}$, which is proportional to the vector spin at the centre of the bond.

The magnetic charge associated with each vertex is given by a discrete, on lattice Gauss' law; $\sum_J M_{IJ} =-Q_I$, where the sum goes over the four nearest neighbours $J$ to site $I$. The minus sign allows for the satisfaction of properties of both the emergent field and the real magnetostatic problem of spin ice \cite{Lhotel2020fragmentation}. The vertex charge takes values $Q_I=0,\pm Q,\pm 2Q$, where $Q=2m/a$ is the monopole charge \cite{Castelnovo2008}.
Labelling the four field elements $[M_{IJ}]$ in order $1\dots 4$ (see eqn. (\ref{vertexorder})), a vertex satisfying the ice rule, with $Q_I=0$ (two spins pointing in and two out) can be written $[M_{IJ}]=[1,1,-1,-1]$ in units of $m/a$. Using the same notation, monopole carrying vertices are of the form $[M_{IJ}]=\pm[1,-1,-1,-1]$ for $Q_I=\pm Q$.

At this microscopic level, the Helmholtz-Hodge decomposition implies that each vertex set  $[M_{IJ}]$ is cut into three distinct parts indicated by eqn. (\ref{Helmholtz})
\begin{equation}
    [M_{IJ}]=[M_{IJ}]_{\vb m}+[M_{IJ}]_{\vb d}+[M_{IJ}]_{\vb h}.
\end{equation}
Here $\vb m$ stands for monopole and represents the divergence full longitudinal part, $\vb d$ the divergence free transverse part and $\vb h$ the harmonic contribution. 

The decomposition can be calculated for any spin configuration by first identifying the vertices carrying magnetic charge and solving for the longitudinal field components via the Poisson equation \cite{Faulkner2015}. Assuming periodic boundary conditions, the sum of the transverse and harmonic contributions is then the difference, $[M_{IJ}]-[M_{IJ}]_{\vb m}$, which should satisfy Kirchoff's current law at each vertex. A more practical alternative  method \cite{Slobinsky2019}, iteratively calculates the divergence free part $[M_{IJ}]_{\vb d}+[M_{IJ}]_{\vb h}$ for a given charge distribution, yielding $[M_{IJ}]_{\vb m}$ by the appropriate subtraction. 

The harmonic contribution can be understood by considering the solution to Poisson's equation for charges distributed on a torus. It is invariant on adding a term $\psi'(\vec r)=\vec h.\vec r$ to the scalar potential, with $\vec h$ the spatially uniform harmonic field. As a consequence, multi-valued solutions are analytically connected by winding a charged particle around the torus, returning to its starting position \cite{Maggs2002,Faulkner2015}. Winding a charge $q$ along the $\hat{z}$ axis of a torus of scale $L$ in dimension $d$ would change $\vec h$ by  $\delta \vec h\sim \frac{q}{L^{d-1}} \hat{z}$.

Moving these arguments directly to spin ice puts us on the diamond lattice of charge vertices with the cubic axes lying along the principle directions of the torus. The individual solutions correspond to different topological sectors \cite{Jaubert2013} which fix the topological contribution to the magnetisation. For simplicity here we consider a situation with monopole concentration zero and magnetisation maintained along the $[001]$ direction either by an external field or by a symmetry breaking perturbation \cite{Jaubert2010}. The average magnetisation per spin is then $\vec M = \frac{m}{\sqrt{3}}\epsilon \hat{z}$ from which we can identify a harmonic fragment for each field element of amplitude $\epsilon$ in units of $m/a$. For any vertex lying on the $A$ sublattice, the harmonic flux flows out along elements $1$ and $4$ and in through $2$ and $3$ (see eqn.(\ref{vertexorder})):
\begin{equation}
[M_{IJ}]_h=\frac{m}{a}[\epsilon,-\epsilon,-\epsilon,\epsilon], \; 0\le \epsilon\le 1,
\end{equation}
while for a $B$ sub-lattice the signs are reversed. As the magnetisation becomes saturated, $\epsilon \rightarrow 1$ and the harmonic fragment takes on $100\%$ of the magnetic resources. 

The topological harmonic fragment remains defined even in the presence of a finite monopole concentration \cite{Faulkner2015}. In this case, it will be dressed by a paramagnetic contribution to the magnetisation due to the statistics of monopole configurations of finite extent. We choose to include this contribution as part of the longitudinal fragment, but both ultimately contribute to the magnetisation and its fluctuations \cite{Jaubert2013}. More realistic boundaries, with fixed surface charges and defects \cite{Revell2012} will result in a harmonic component with some structure. Consequently this will generate some diffuse scattering at finite $\vec q$ in addition to the topological contribution at $\vec q=0$.

The topological sectors are iso-energetic (unlike for a standard fluid of electric charges \cite{Faulkner2015}), but in zero field the sector straddling zero magnetisation is selected entropically.
As a consequence, in the monopole fluid phase, in zero external field, the harmonic contribution is zero to a good approximation, so that the field built from the magnetic moments decomposes into two ``orthogonal'' fluids with elements  $[M_{IJ}]_{\vb m}$ and  $[M_{IJ}]_{\vb d}$ \cite{Brooks2014}. 

As the monopole concentration goes to zero, only the transverse fragment survives, $[M_{IJ}]\rightarrow [M_{IJ}]_d$, while crossing the phase boundary into the all-in all-out antiferromagnetic phase (a double monopole crystal \cite{Borzi2014,raban2019multiple}), the field elements are purely longitudinal, $[M_{IJ}]\rightarrow [M_{IJ}]_m$. The monopole crystal phase \cite{Brooks2014,Borzi2013,Lefrancois2017,Cathelin2020,Pearce2022} is intermediate between these two limits \cite{raban2019multiple} with $[M_{IJ}]$ divided evenly between $[M_{IJ}]_{\vb m}$ and  $[M_{IJ}]_{\vb d}$. The monopole part forms long range all-in-all-out order and the dipolar part a Coulomb liquid with characteristic dipolar correlations.

\section{From the Kagom\'e Plateau of Spin Ice to Kagom\'e Ice} 
\label{plateau-ice}

\begin{figure}[t]
    \centering
    \includegraphics[width=0.7\linewidth]{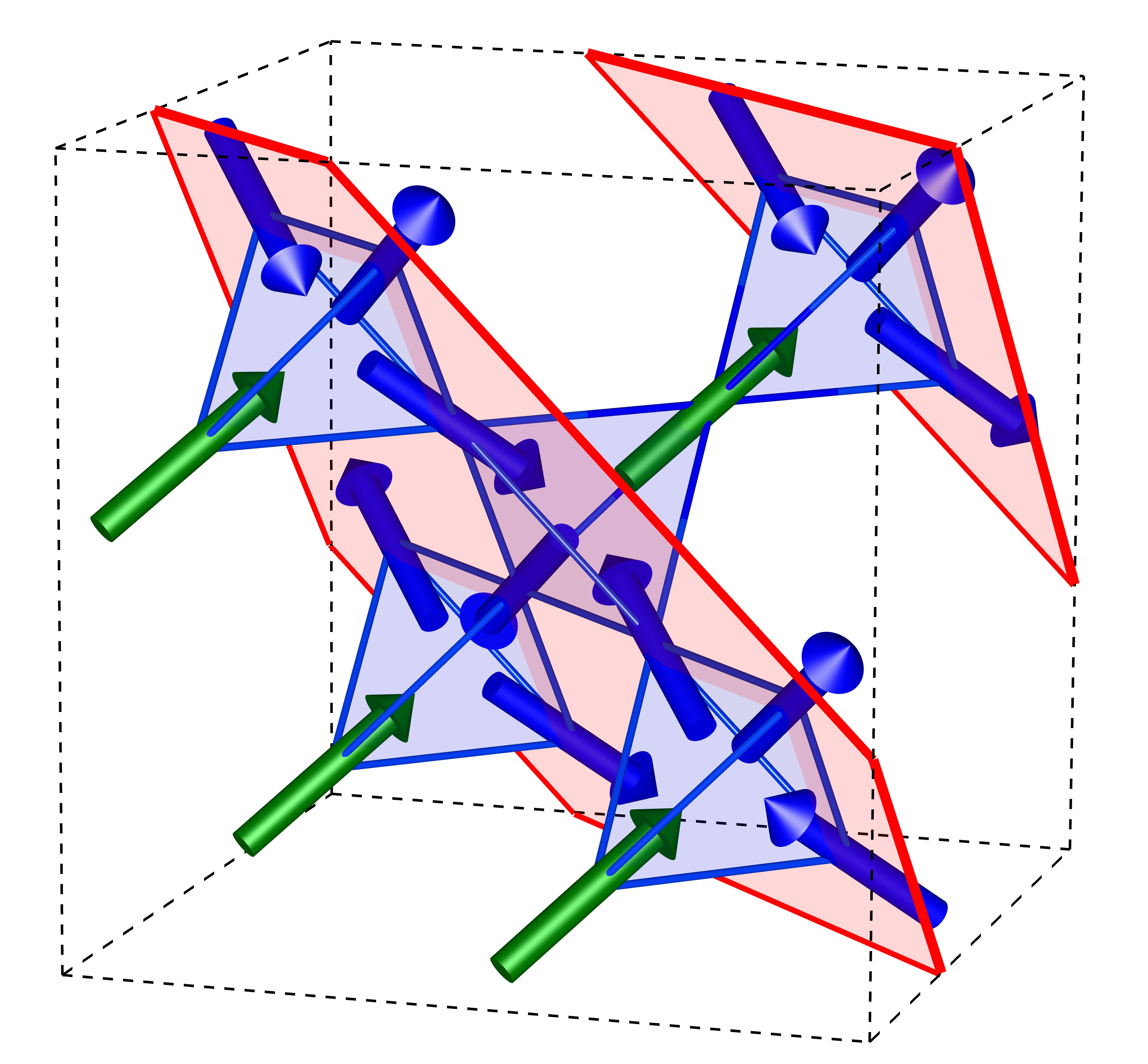}
    \caption{Pyrochlore spin ice in a [111] field, showing the distinction between planes of pinned apical spins on a triangular lattice (green) and kagome planes satisfying the kagome ice rules (red).}
    \label{pyrochlore-111fragmentation}
\end{figure}

Applying a magnetic field of modest strength along the $[111]$ body centred cubic axis aligns the apical spins of each tetrahedron along the field direction, as shown in Fig. (\ref{pyrochlore-111fragmentation}). As the monopole concentration goes to zero, the system enters the kagom\'e plateau region \cite{Isakov2004b} in which kagom\'e planes of spins lying perpendicular to the field direction enter the KII topological liquid phase with residual entropy at low temperature. In each tetrahedron the ice rules of two spins in and two out are satisfied but as the apical spin is fixed to be out for an $A$ tetrahedron and in for $B$, the three remaining spins in the in-plane triangles satisfy the kagom\'e ice rule with two spins in and one out on an $A$ triangle and two out one in on a $B$ triangle \cite{Moessner2003}. 

This evolution is well captured by fragmentation. A magnetic moment along the [111] axis can be decomposed into three cubic contributions of equal amplitude, each of which generates an independent harmonic fragment. Following eqn. (\ref{vertexorder}), a vertex of type $A$ has harmonic fragment
\begin{eqnarray}
[M_{IJ}]_h &=& [M_{IJ}]_h^x+[M_{IJ}]_h^y+[M_{IJ}]_h^z  \\ \nonumber
[M_{IJ}]_h &=& [-\epsilon,\epsilon,-\epsilon,\epsilon]+[-\epsilon,-\epsilon,\epsilon,\epsilon]+[\epsilon,-\epsilon,-\epsilon,\epsilon]\\ \nonumber
[M_{IJ}]_h &=& [-\epsilon,-\epsilon,-\epsilon,3\epsilon], \; 0\le \epsilon\le \frac{1}{3}.
\end{eqnarray}
The kagome plateau corresponds to $\epsilon=\frac{1}{3}$ so that for one of the three vertex configurations with spin $4$ pointing along $[111]$
\begin{align}
    [M_{IJ}] & = [-1,-1,1,1] \nonumber                                                                                                      \\ \label{F111}
           & = [0]_{\vb m} + [-\frac{2}{3}, -\frac{2}{3}, \frac{4}{3}, 0]_{\vb d} + [-\frac{1}{3}, -\frac{1}{3}, -\frac{1}{3}, 1]_{\vb h}.
\end{align}
The longitudinal fragment is zero, the transverse fragment is restricted to the three spins in the plane with two elements of amplitude $2/3$ and one of $4/3$ which together satisfy Kirchoff's law. The harmonic term is identical for each tetrahedron or unit cell,  spreading out evenly over the three in-plane spins. The apical spin is purely harmonic and  the sum over the contributions also satisfies the current law.

Spin ice fragmentation on the kagom\'e  plateau is intimately related to the fragmentation of two-dimensional spins in kagom\'e ice. In this case the basic spin units are triangles whose centres form a honeycomb lattice of vertices for magnetic charge accumulation \cite{Moller2009,Chern2011,Brooks2014,Canals16}. Considering the spins in an isolated kagom\'e layer on the kagom\'e plateau, the in-plane projection of the harmonic terms leaves a magnetic charge accumulation at the honeycomb lattice sites corresponding to the magnetic charge crystal observed in the KII  phase of kagom\'e ice \cite{Moller2009,Chern2011}. The three-dimensional harmonic term therefore corresponds to a two dimensional longitudinal term.  
Using a similar notation to above the three, two-dimensional field elements entering a triangle of type $A$ can be written
\begin{align}
    [M_{IJ}]^{2D} & = [-1,-1,1] \nonumber     \label{Fkag}                                                                                             \\
                  & = [-\frac{1}{3}, -\frac{1}{3}, -\frac{1}{3}]_{\vb m} + [-\frac{2}{3}, -\frac{2}{3}, \frac{4}{3}]_{\vb d} +[0]_{\vb h}.
\end{align}
The units of the field elements are $\frac{2m}{a\sqrt{2}/3}$ accommodating the projection of the three dimensional spin vectors onto the plane \cite{Turrini2022} and the charge accumulation at the honeycomb vertices is only one half of the in plane monopole charge \cite{Castelnovo2008,Fulde2002}.
An example of such a decomposition is shown in Fig. (\ref{starphase-frag}) for the ordered $\sqrt{3}\times\sqrt{3}$ phase discussed in more detail in the next section.

\section{Neutron Scattering from Fragmented States}
\begin{figure*}
    \centering
    \includegraphics[width = 0.95\linewidth]{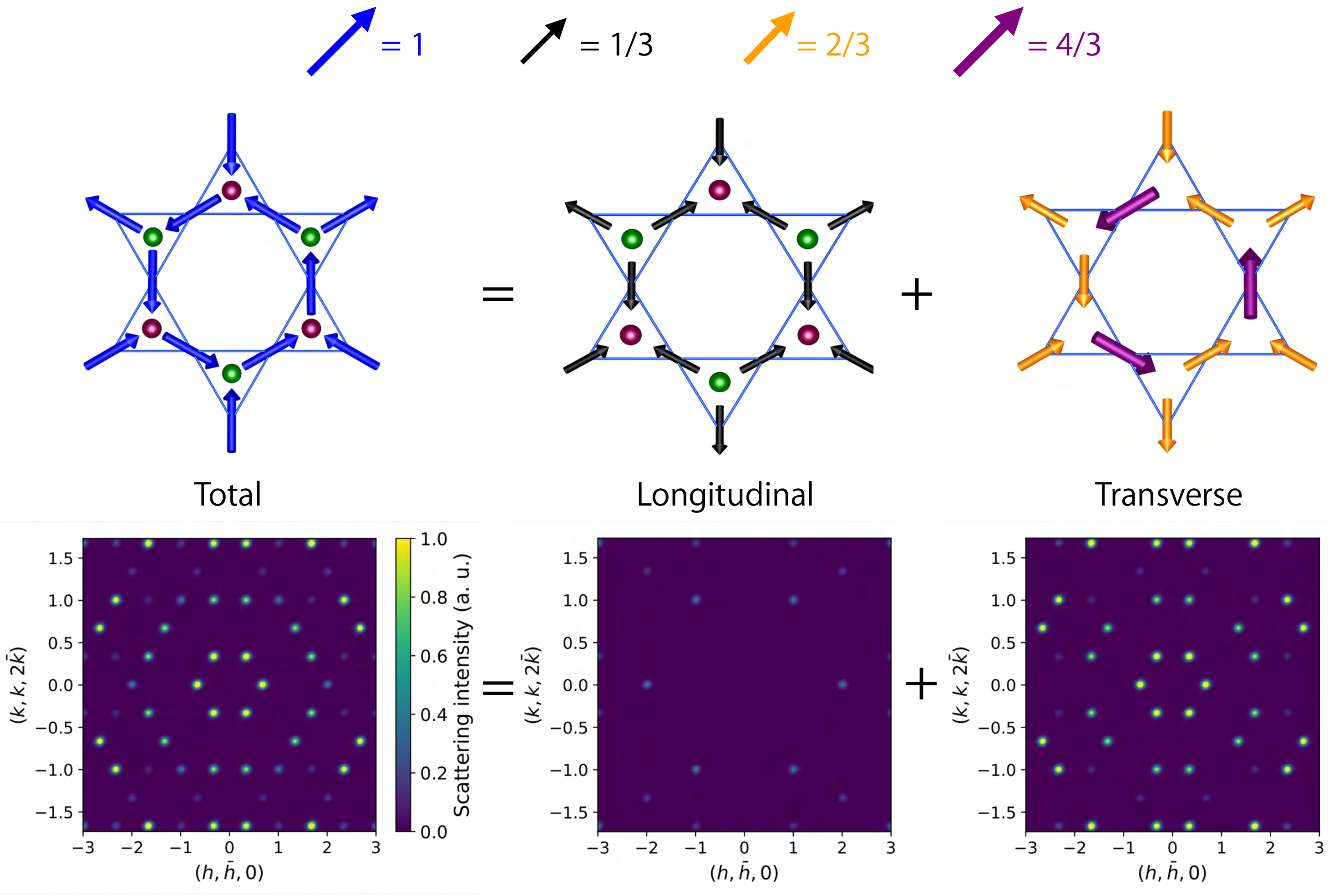}
    \caption{Top: Fragmentation of the $\sqrt{3}\times\sqrt{3}$ phase magnetic structure on a kagom\'e plane. The magnetic unit cell extends over 9 sites. Colours illustrate the amplitude of each component, and green and purple spheres show the placement of positive and negative magnetic charges within the dumbbell model. Left panel, full spin configuration.  Middle panel, longitudinal fragment $\vec M_{\vb m}$ showing ``all spins in all spin out'' ordering. Right panel, transverse fragment $\vec M_{\vb d}$ showing emergent ordering of the star phase. Bottom: SF neutron scattering intensities for neutrons polarised perpendicular to the plane computed from the total and corresponding fragment above. The total scattering picture can also be computed by adding the separate intensities of the two fragments.}
    \label{starphase-frag}
\end{figure*}

The fragmentation decomposition is particularly useful for magnetic neutron scattering as the longitudinal and transverse fragments, when transformed into reciprocal space, are mutually orthogonal, while the topological harmonic fragment is restricted to wave vector $\vec q=\vec 0$ and subsequent Brillouin zone centres. For a system of $N$ spins the Fourier transform of a magnetic configuration is defined
\begin{align}
    \vec M(\vec q) & = m\sum_{i=1,N} \vec S_i \exp(i\vec q.\vec r_i) \nonumber                           \\
                   & = a\sum_{I=1,N/4}\sum_{J=1,4} \vec M_{IJ} \exp(i\vec q.(\vec r_I + \vec \delta_J)),
\end{align}
The spin and tetrahedron centres are at positions $\vec r_i$ and $\vec r_I$ respectively and $\vec \delta_J= \frac{a}{2}\vec d_J$.

Following the spin fragmentation we can write $\vec M(\vec q)=\vec M(\vec q)_{\vb m}+ \vec M(\vec q)_{\vb d}+ \vec M(\vec q)_{\vb h}$.
The component $\vec M(\vec q)_{\vb d}$ is ``transverse'' in that it lies perpendicular to the wave vector $\vec q_{\ast}=\vec q - \vec G$, which is folded back into the first Brillouin zone by the appropriate reciprocal lattice vector $\vec G$. Both the ``longitudinal'' component, $\vec M(\vec q)_{\vb m}$ and the harmonic component $\vec M(\vec q)_{\vb h}$, lie parallel to $\vec q_{\ast}$ with the latter restricted to the Brillouin zone centres.

Neutron scattering within the static approximation gives access to the Fourier transform of the two site, one time, spin-spin correlation function
\begin{equation}
    S^{\alpha\beta}(\vec q)=\left<M_\alpha(\vec q)M_{\beta}(-\vec q)\right>,
\end{equation}
where $\alpha$, $\beta$ are cartesian indices $x,y,z$, $\vec q$ is the wave vector transfer of the scattering process and $\left<\dots\right>$ represents a thermal average.
The neutron scattering cross section is proportional to the projection of the correlation tensor perpendicular to $\vec q$
\begin{equation}
    S(\vec q)=\left<|\vec M_\perp(\vec q)|^2\right>,
\end{equation}
where $ \vec M_{\perp}$ is the projection of $\vec M$ perpendicular to $\vec q$.
For simplicity we take the magnetic form factor to be a constant, independently of $\vec q$.

As a consequence of the orthogonality condition  the structure factor also decomposes into distinct parts
\begin{align}
    S(\vec q) & = S(\vec q)_{\vb m}+S_{\vb d}(\vec q),
\end{align}
so that the scattering intensity divides into components from the  divergence full (plus harmonic at the zone centres) and divergence free fragments of the magnetic moments with no interference terms. In the following sections we will demonstrate this property by computing the elastic scattering intensity of each fragment as well as of the total spin structure for different fragmented magnetic  states.
This property opens up the possibility of defining fragmentation order parameters by integrating the scattered intensity in specific regions of reciprocal space.

Inside the first Brillouin zone the scattering is purely transverse: $S(\vec q)=S_{\vb d}(\vec q)$. For larger $\vec q$, as the scattering cross section lies perpendicular to $\vec q$ rather than $\vec q_{\ast}$, $S(\vec q)$ develops contributions from the other two fragments. 
The separation of these fragments has already been observed in magnetic charge crystal phases \cite{Brooks2014,Lefrancois2017,Cathelin2020,Canals16,Paddison2016}.  In these phases the harmonic component can be ignored,  the longitudinal fragment gives antiferromagnetic long range order corresponding to the ordered array of magnetic charges and the transverse part gives diffuse scattering characteristic of the Coulomb spin liquid phase \cite{Henley2010}.

In the case of polarised neutrons, $S(\vec q)$ can be further resolved into ``spin flip'' (SF) and ``non-spin flip'' (NSF) components corresponding to scattering events in which the neutron spin direction is flipped or not \cite{Fennell2009}. The SF scattering cross section lies in the plane perpendicular to the polarisation axis and  projects out the component of  $\vec M_{\perp}(\vec q)$ lying in this plane. The NSF component projects $\vec M_{\perp}(\vec q)$ onto the polarisation axis. This refinement leads to separate contributions to the structure factor,
$S(\vec q)^{SF}$ and $S(\vec q)^{NSF}$ for scattering perpendicular and parallel  to the polarisation axis, each of which can be decomposed into the perpendicular fragmentation components. For an unpolarised source the measurement averages over all polarisation directions leaving the total scattering intensity proportional to $S(\vec q)$. Polarised neutron refinement is of particular interest for scattering from spin ice materials on the kagom\'e  plateau. In this case, choosing the neutron polarisation along the $[111]$ field direction allows for the resolution of spin components parallel and perpendicular to the kagom\'e plane \cite{Turrini2022}.

As a specific example we show the decomposition of the scattering intensity from a two-dimensional sample of kagom\'e ice.
A possible evolution of the classical KII phase as the temperature is lowered, is to the ``$\sqrt{3}\times\sqrt{3}$'' phase whose structure is illustrated in Fig.~(\ref{starphase-frag}) top \cite{Moller2009,Chern2011}. The repetition distance for the unit cell is $\sqrt{3}$ larger than that of the kagom\'e lattice. As shown, the spins in the unit cell can be fragmented. The longitudinal part gives the charge order of alternate positive and negative charges, with a reduced unit cell of three sites.
The transverse part maintains the 9-site unit cell, whose configuration maps onto the emergent field of a dimer solid, the ``star phase'' in which a tiling of the unit cells produces three distinct types of hexagonal ring. One out of the three types of hexagon forms a six-fold symmetric star of dimers from which the phase takes its name \cite{Moessner2001,Schlittler2017}.
The longitudinal, transverse and total contributions to the scattering intensity from this ordered state are shown in Fig.~(\ref{starphase-frag}) bottom.
For the kagom\'e plateau of spin ice, this in-plane scattering intensity would correspond to $S(\vec q)^{SF}$ with the neutron source polarised along the $[111]$ direction. In this case the longitudinal fragment corresponds to the projection onto the plane of the three dimensional harmonic component, which is channelled out of each tetrahedron via the apical spin (not included) \cite{Turrini2022}  (see eqns. (\ref{F111}) and (\ref{Fkag})).
The data is shown in the scattering plane of the kagom\'e lattice in units appropriate for spin ice and the kagom\'e plateau: the in-plane axes $[k,k,2\bar{k}]$, $[h,\bar{h},0]$ lie perpendicular to the $[111]$ field axis and are in units of $2\pi/a_c$. The six fold symmetry of the spins lying in the plane is represented in the figure by scaling the $[k,k,2\bar{k}]$ axis by a factor of $\frac{1}{\sqrt{3}}$.

This analysis shows that, rather surprisingly this simplest of phases, the classical $\sqrt{3}\times\sqrt{3}$ phase is a fragmented double-$q$ structure whose scattering pattern is the sum of intensities from the longitudinal and transverse parts. These fragments have different ordering wave vectors and have no communal Bragg peaks so that the total scattering is made up of resolvable contributions from the charge ordering and the emergent field from the star phase.
The charge ordering from the longitudinal component corresponds to antiferromagnetic, ``all spins in all spins out'' order. This is a ``$\vec {q}=0$'' order, with Bragg peaks at the centres of the kagom\'e lattice Brillouin zone starting at $h=2,k=0$ and symmetry related points, the scattering intensity being zero at the zone centres with smaller wave vector transfer. The star phase from the transverse component shows Bragg peaks at $h=\frac{2}{3}$, $k=0$ and symmetry related points. These correspond to the basis vectors of the reciprocal space for the $\sqrt{3}\times\sqrt{3}$ unit cell  with magnitude $q= \frac{2\pi}{a_c}\left(\frac{2\sqrt{3}}{3}\right)$. Peaks at larger $q$ repeat in a distinctive, 6-fold symmetric pinwheel pattern which we can take to be characteristic of the star phase.

\begin{figure*}
    \centering
    \includegraphics[width = 0.95\linewidth]{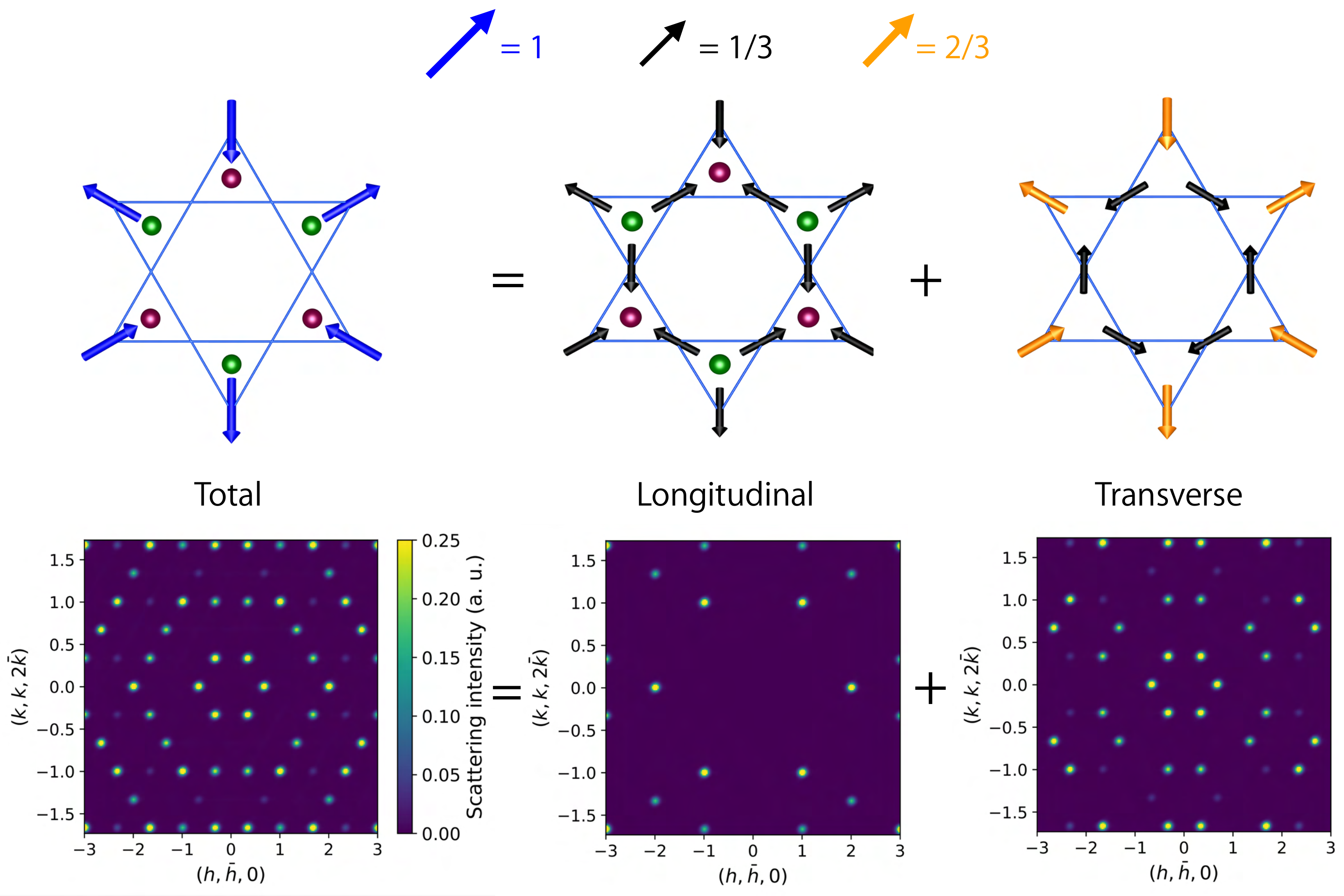}
    \caption{Top: Fragmentation of the spin-$P$ phase magnetic structure on a kagom\'e plane. The magnetic unit cell extends over 9 sites. Colours illustrate the amplitude of each component, and green and purple spheres show the placement of positive and negative magnetic charges within the dumbbell model. Left panel, full spin configuration. The quantum resonance on the hexagonal loop results in the effective absence of spins around the loop. Middle panel, longitudinal fragment $\vec M_{\vb m}$ showing ``all spins in all spins out'' ordering. Right panel, residual transverse fragment $\vec M_{\vb d}$ corresponding to the residual emergent field of the dimer plaquette phase (see Fig. \ref{superp-plaq-trans} top). Bottom: SF neutron scattering intensities for neutrons polarised perpendicular to the plane computed from the total and corresponding fragment above. The total scattering picture can be computed by adding the separate intensities of the two fragments. Note that the absolute intensity scale is one quarter of that in Fig~(\ref{starphase-frag}).}
    \label{plaqphase-frag}
\end{figure*}

\section{Quantum Fluctuations: the spin-$P$ and Plaquette Phases}
In this section we consider the effect of quantum fluctuations on the KII phase of kagom\'e ice. It is known that quantum fluctuations driven by a small transverse spin component could drive the spins into a partially ordered phase at low temperature \cite{Bojesen2017,Wu2019,Wang2020}. In this resonating $\sqrt{3}\times\sqrt{3}$ phase which we refer to as the spin-$P$ phase, two of the three types of hexagonal spin arrangement provide a framework for resonating loops of  six spins around the third class of hexagon. This quantum resonance corresponds to a linear superposition of the two states per unit cell with spin rotations around the enclosed hexagon in opposite directions, leaving an effective magnetic state with reduced total moment, as shown in Fig. (\ref{plaqphase-frag}) top left.

\begin{figure}
    \centering
    \includegraphics[width = 1\linewidth]{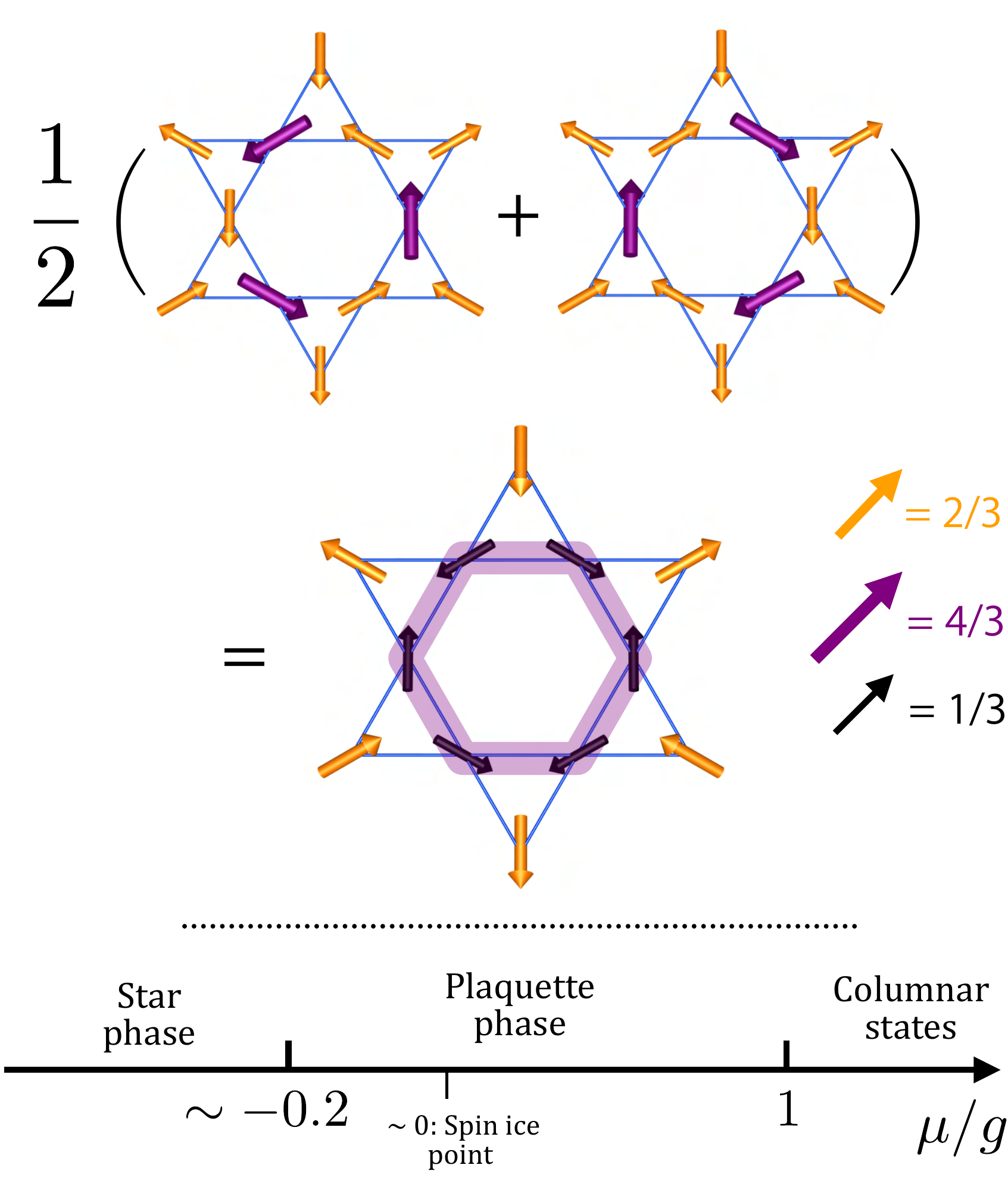}
    \caption{Top: the quantum resonance of the emergent field for dimers in the plaquette  phase is modelled as the average of left and right circulations. Emergent dimers are located on the purple minority spins. The resultant quantum superposition is shown below, with a purple shade illustrating the dimer resonance of the plaquette phase. Colours illustrate the amplitude of each spin component. Bottom: phase diagram of the dimer model given by eqn. (\ref{hex-flip}) on the honeycomb lattice \cite{Moessner2001}.}
    \label{superp-plaq-trans}
\end{figure}

To show that the spin-$P$ phase corresponds to the coexistence of the classical charge ordered phase and an emergent quantum dimer phase \cite{Moessner2001,Schlittler2017} one must first apply the fragmentation procedure to the effective reduced moments once the quantum spin resonances have been taken into account. From Fig. (\ref{plaqphase-frag}) top, one can see that the residual spin on each triangle can be written, using the previous notation;  $\pm [-1,0,0]$, arranged such that the charge order is preserved. A vertex carrying a positive charge can thus be fragmented into a longitudinal and a transverse part
\begin{equation}
    [-1,0,0] = [-\frac{1}{3}, -\frac{1}{3}, -\frac{1}{3}]_{\vb m} + [-\frac{2}{3}, \frac{1}{3}, \frac{1}{3}]_{\vb d} +[0]_{\vb h} \label{P-frag}\, ,
\end{equation}
This decomposition confirms that, on driving the $\sqrt{3}\times\sqrt{3}$ phase into the spin-$P$ phase with quantum fluctuations, the charge ordering and hence the longitudinal fields are unchanged, while the amplitude of the transverse part is reduced by a factor of two. The quantum resonance is therefore limited to the transverse fragment as announced.
In dimer language, adding quantum fluctuations to the star phase leads to resonating closed loops of dimers which can lead to a quantum phase transition to the ``plaquette phase''. This is not a liquid phase, as dimer translational symmetry remains broken such that resonances are limited to plaquette flips of dimers around one of the three types of hexagon of the star phase. The corresponding resonance of the emergent field for the dimers is shown in Fig. (\ref{superp-plaq-trans}) top. Similarly to the residual spin of the spin-$P$ phase, it is constructed as the average of both emergent field configurations around a plaquette. Despite the resonance, the field retains a static residue which is precisely that of transverse spin fragment shown in eqn. (\ref{P-frag}). The spin-$P$ phase is therefore a superposition of the charge ordered phase and the  dimer plaquette phase represented by a single $Z_2$ sector of its emergent field.

In the simplest quantum dimer models, quantum fluctuations are generated through off-diagonal couplings between classical configurations that generate the hexagonal plaquette flips of dimers \cite{Moessner2001,Schlittler2017}. The off-diagonal energy scale, $g$ is in competition with a  classical, three body interaction term, $\mu$ giving an energy scale for a three dimer hexagon and an internal energy for each dimer configuration. An effective Hamiltonian for such a system can be written
\begin{multline}
    \mathcal{H}_{\text{eff}} = \mu \sum_{\hexagon} \left( \ketbra{\hexagona}{\hexagona} + \ketbra{\hexagonb}{\hexagonb} \right)  \\ - g \sum_{\hexagon} \left( \ketbra{\hexagona}{\hexagonb} + \ketbra{\hexagonb}{\hexagona} \label{hex-flip} \right),
\end{multline}
where the sum over $\hexagon$ is over all hexagonal loops containing three dimers, depicted by double links. For $g=0$ and $\mu<0$ one finds the classical star phase dimer solid \cite{Moessner2001,Schlittler2017} and for $\mu>0$, the columnar phase which maps onto the ferromagnetically ordered phase for both kagom\'e and spin ice \cite{Moessner2003}. As $g$ increases from zero, the star phase order parameter is progressively reduced from saturation \cite{Schlittler2017} up to a threshold  and a discontinuous transition into a small window around $\mu/g=0$, in which quantum fluctuations favour the plaquette phase  (See Fig. (\ref{superp-plaq-trans}) bottom). In the mapping between quantum spin ice and dimer problems $\mu<0$ could be thought of as representing the corrections to the dumbbell model from the long range interactions, that are characteristic of spin ice materials \cite{denHertog2000,isakov2005projective} and artificial spin ice \cite{Nisoli2013}  as, in  classical kagom\'e ice they also drive the system into the $\sqrt{3}\times\sqrt{3}$ phase at low temperature \cite{Chern2011}. In the direct mapping from classical nearest neighbour models $\mu$ is zero but it is often left as a renormalizable, free parameter \cite{Benton2012}.

Neutron scattering data from the spin-$P$ phase is easily interpreted using the fragmentation picture. We again expect the data to separate into independent longitudinal and transverse components and predict that the transverse scattering intensity will be reduced by a factor of four compared with scattering from the classical $\sqrt{3}\times\sqrt{3}$ phase. This is confirmed in Fig. (\ref{plaqphase-frag}) where we show calculated neutron scattering data from the spin-$P$ phase for neutrons polarised perpendicular to the scattering plane. The intensity scale is reduced by a factor of four compared with Fig. (\ref{starphase-frag}), highlighting the relative change in the two intensities. The peak structure is identical for the classical and quantum phases but the intensity difference can be used as a diagnostic to distinguish between them. For example, in the classical limit for the $\sqrt{3}\times\sqrt{3}$ phase, the intensity of inner ring of star phase peaks at $h=\frac{2}{3}$, $k=0$ and symmetry related points, $I_d^s$ is four times that of charge ordering peaks at $h=2$, $k=0$ and related points, $I_m$, while in the spin-$P$ phase, the two sets of peaks, $I_d^p$ and $I_m$ have the same intensity. 

This difference
could be used as the basis for an order parameter: 
\begin{equation}
Q=\sqrt{\frac{4I_m-I_d}{3I_m}}, \label{OPQ}
\end{equation}
which distinguishes between the two phases, with $Q=0$ for the classical $\sqrt{3}\times\sqrt{3}$ ground state and $Q=1$ for a perfect plaquette phase. This order parameter has the advantage over the one used in quantum Monte Carlo simulations of dimers  \cite{Schlittler2017}, of being built from experimental observables. 

The most recent numerical results suggest that the star to plaquette quantum phase transition is first order \cite{Schlittler2017}. As a consequence  we anticipate that  $Q$ will undergo a discontinuous jump at the transition.

\section{Quantum Spin Ice in a $[111]$ Field}

\begin{figure}
    \centering
    \includegraphics[width = 1\linewidth]{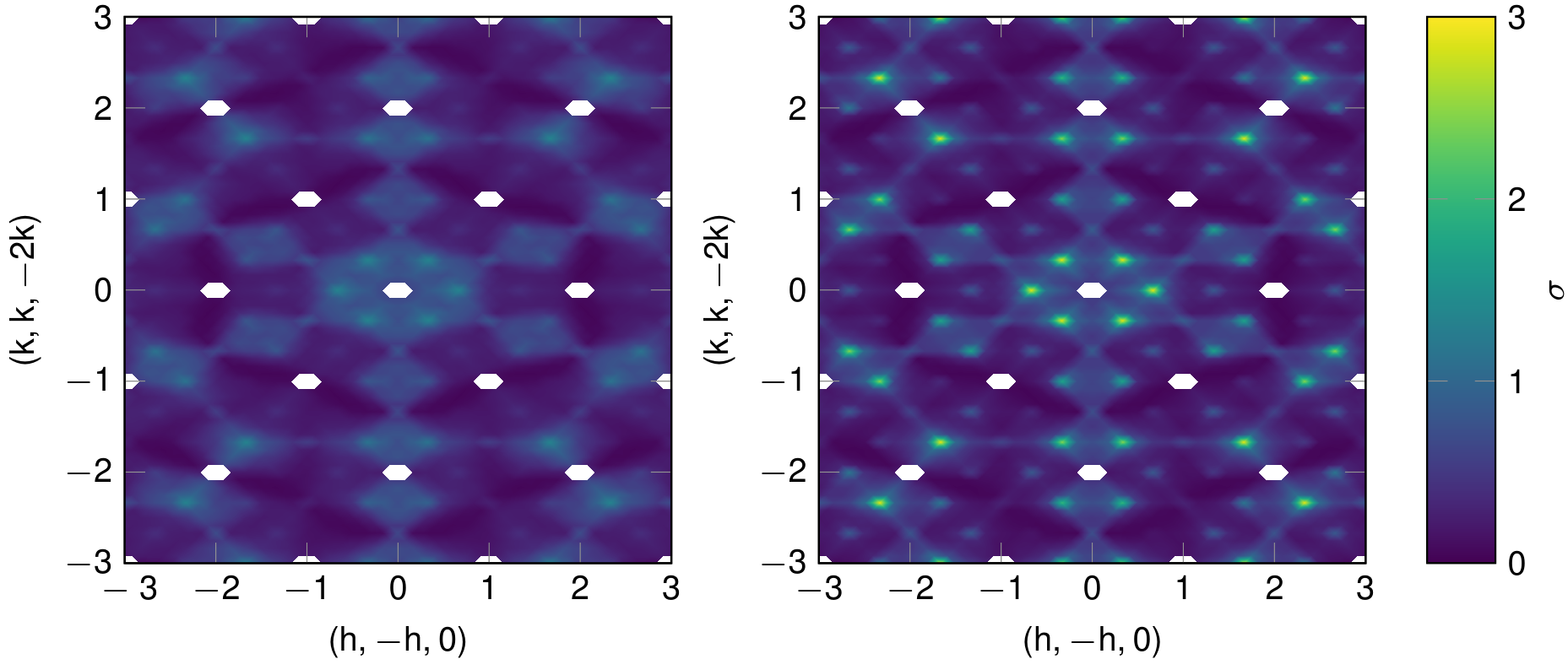}
    \caption{Unpolarized neutron scattering data in the kagom\'e plane from Quantum Monte Carlo simulations of quantum nearest neighbour spin ice in a $[111]$ field. Left panel $\frac{T}{J}=\frac{1}{20}$, right panel  $\frac{T}{J}=\frac{1}{320}$. Data reproduced with permission from \cite{Bojesen2017}.}
    \label{Onoda}
\end{figure}

In this section we review data from existing work in the context of the fragmentation picture.
Shown in Fig. (\ref{Onoda}) is constructed unpolarised neutron scattering data in the kagom\'e plane from Quantum Monte Carlo simulations. The data, for nearest neighbour quantum spin ice in a $[111]$ field is reproduced from Ref. [\onlinecite{Bojesen2017}]. It is taken in the intermediate field region corresponding to the kagom\'e plateau. The right hand panel shows data taken at $\frac{T}{J}=\frac{1}{20}$ where $T$ is the temperature and $J$ the coupling constant. It is consistent with Coulomb phase spin liquid behaviour, showing correlated diffuse scattering with pinch point features \cite{Moessner2003,Turrini2022}. Bragg peaks at the Brillouin zone centres 
($h=2$, $k=0$ for example) are masked. In an experiment these would also coincide with the structural Bragg peaks.  On the right we show data at much lower temperature, $\frac{T}{J}=\frac{1}{320}$ where the development of order is clearly observed. A different choice of scale along the vertical axis distorts the 6--fold symmetry of the scattering pattern but despite this one can observe features similar to those shown in Figs. (\ref{starphase-frag}) and  (\ref{plaqphase-frag}). In particular, sharpening peaks at $h=\frac{2}{3}$, $k=0$ and symmetry related points are clearly visible at the lower temperature. These are the first elements of the radial pattern of peaks characteristic of the pin-wheel ordering of the emergent dimers. Moving out along one of the spokes of the pattern the characteristic alternation of high and low intensity peaks is visible. Even here a diffuse scattering background remains, due presumably to remnant incoherent or thermal spin fluctuations about the ordered phase.
The figures also show additional peaks compared with Figs. (\ref{starphase-frag}) and  (\ref{plaqphase-frag}). These are due to scattering from the out of plane spin components which appear as a consequence of simulating an unpolarised neutron source.

The order parameter $Q$, (eqn (\ref{OPQ})) could be used as a diagnostic tool to distinguish between the classical  $\sqrt{3}\times\sqrt{3}$ and quantum spin-$P$ phases.  For this one would need to include and analyse the magnetic peak intensities at the zone centres coming from the harmonic fragment. Projections of the harmonic sector parallel and perpendicular to the kagom\'e planes provide both the three dimensional ferromagnetic order and the two-dimensional charge order which can be separated using the analysis of section \ref{plateau-ice}. 
In experiment the total intensity at these points also includes a dominant structural contribution. An independent estimate of this intensity is necessary. This would be subtracted from the total to give the magnetic scattering intensity. However, in the present case of numerical simulation, the simulated intensities could be compared directly with predicted values.

\section{Neutron Scattering from the Monopole Crystal Phase of spin ice}

Similar logic involving the fragmentation protocol can be applied to the ordering process out of the partially ordered  fragmented monopole crystal phase of spin ice in zero magnetic field \cite{Brooks2014,Jaubert2015} and this is the subject of this section. In the monopole crystal phase a vertex carrying a south pole (negative charge) takes the form
\begin{align}
    M_{IJ} & = [1,1,1,-1]                                                                                                                         \\
           & = [\frac{1}{2},\frac{1}{2},\frac{1}{2},\frac{1}{2}]_{\vb m} + [\frac{1}{2}, \frac{1}{2}, \frac{1}{2},-\frac{3}{2}]_{\vb d},\nonumber
\end{align}
and the transverse fragment maps onto one of the $Z_2$ sectors of the emergent field for hard core dimers on a diamond lattice \cite{Huse2003}. In this case, the element carrying the flux of magnitude $\left(\frac{m}{a}\right)\left(\frac{3}{2}\right)$, which is the minority spin of either the ``three in one out'' or the ``three out one in vertex'' corresponds to the dimer position. Monopole charge ordering again coexists with an effective classical dimer liquid represented by the transverse spin fragment.

\begin{figure*}[t]
    \centering
    \includegraphics[width = 0.95\linewidth]{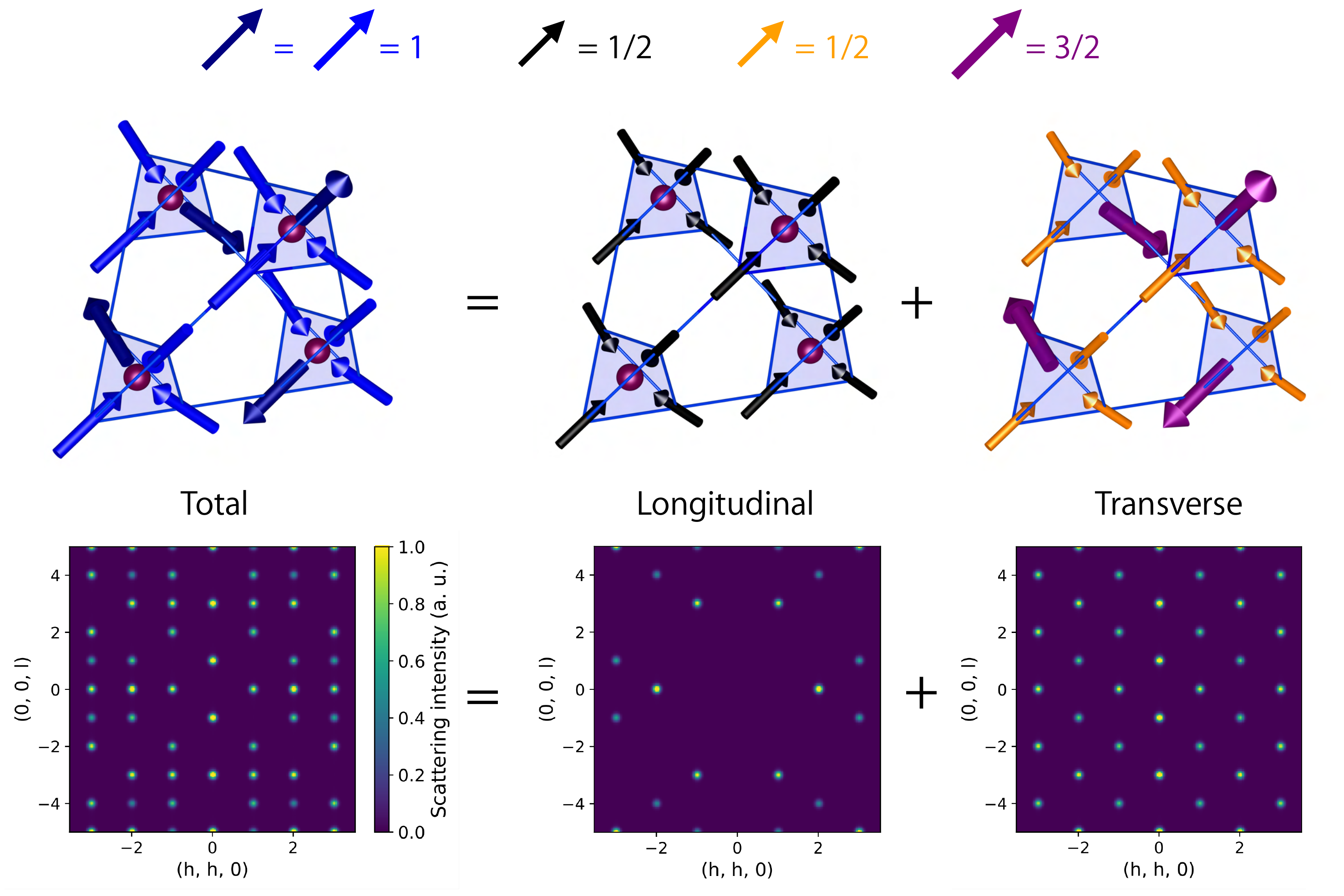}
    \caption{Top: fragmentation of the spin-$R$ phase magnetic structure - a monopole crystal with ordered transverse fragment. Only half of the tetrahedra are pictured for clarity. Left panel, spin configuration. The minority spins are indicated by a darker shade of blue. Middle, longitudinal fragment $\vec M_{\vb m}$ showing ``all spins in all spins out'' ordering. Right panel, transverse fragment $\vec M_{\vb d}$ corresponding to the emergent field for the ordered dimer phase (the $R$-phase, see Fig. (\ref{R-phase}) top). The colours illustrate the amplitude of each spin component. Bottom: unpolarized neutron scattering intensities in the $[hh0]$, $[00l]$ plane computed from the corresponding fragment above. The total scattering picture can also be computed by adding the separate intensities of the two fragments.}
    \label{strucfact-Rphase}
\end{figure*}

Complete ordering can again be either induced by quantum fluctuations or by classical corrections to the monopole picture.
As in the case of kagom\'e ice, adding corrections to the classical monopole picture through the use of the dipolar spin ice Hamiltonian drives the system into the fully ordered phase illustrated in Fig. (\ref{strucfact-Rphase}) upper panel \cite{Jaubert2015}, which we refer to as the spin-$R$ phase. The upper central and right panels show the longitudinal and transverse fragments respectively.
They show that this can be represented as a classical superposition of the ``all spins in all spins out'' order from the charges and the emergent field from the phase of ordered dimers, the $R$-phase \cite{Sikora2011} and that these distinct phases emerge from the two orthogonal spin fragments.
The 16--fold degeneracy of the spin-$R$ phase can be divided into two sets of $8$ states corresponding to the degeneracy of the $R$-phase. The two sets have reversed monopole ordering on the two sublattices of diamond lattice, each of which is tied to a $Z_2$ sector of emergent dimer field.

The calculated unpolarised neutron scattering intensity from the spin-$R$ phase is shown in the lower panels of  Fig.~(\ref{strucfact-Rphase}) for the $[hh0]$, $[00l]$ plane. They confirm that  the scattering decomposes into a fragmented double-$q$ structure with different ordering wave vectors for the longitudinal and transverse parts. The longitudinal fragment shows the characteristic $q=0$ ordering of the ionic crystal, while the transverse part orders with ${\vec q}=[hhl]$ in units of the reciprocal cubic cell, $\frac{2\pi}{a_c}$ and with $h+l$ an odd number. The total intensity is again built of the two independent fragments with no interference terms.

\begin{figure}
    \centering
    \includegraphics[width = 1\linewidth]{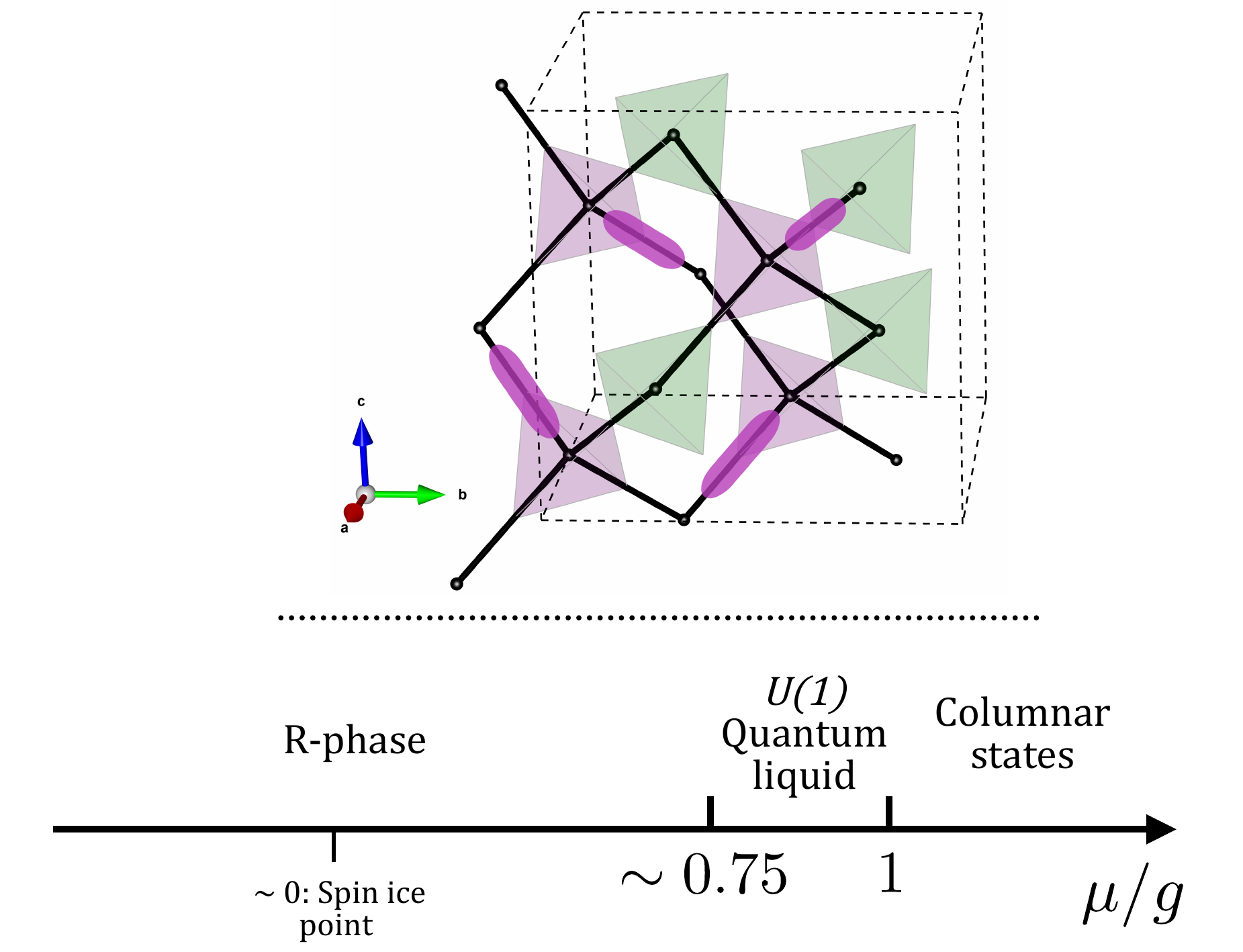}
    \caption{ Top: $R$-phase dimer structure on the diamond lattice. The tetrahedra shown in Fig. (\ref{strucfact-Rphase}) for the   spin-$R$ structure are shaded in mauve. The dimer representation is equivalent to the emergent field representation shown in the top right panel of Fig. (\ref{strucfact-Rphase}). Bottom: phase diagram \cite{Sikora2011} for dimers on a diamond lattice as a function of the ratio $\mu / g$ from eqn. (\ref{hex-flip}). Also shown is the ``spin ice'' point corresponding to the location of the monopole crystal  plus small transverse quantum spin fluctuations \cite{Bergman2006}, deep within the spin-$R$ phase.}
    \label{R-phase}
\end{figure}

Quantum fluctuations can be added to the dimer model via ring exchange flips around closed hexagons for which Hamiltonian (\ref{hex-flip}) can be adapted. This model has been studied both analytically \cite{Moessner2001,Bergman2006} and numerically \cite{Sikora2011}. For large and negative $\mu$, the dimers crystallise into the classical $R$-phase  which maximises the number of hexagonal loops or plaquettes of dimers (Fig. (\ref{strucfact-Rphase}) top left). Switching on the off diagonal term through finite $g$, the system is driven through a quantum phase transition. In this case the transition is to a quantum dimer liquid rather than to a resonating dimer solid. For $\mu/g=1$, hexagonal plaquettes become unfavourable and the system passes discontinuously into a columnar phase {\footnote{referred to as isolated states in \cite{Sikora2011}}}
with dimers aligned along one of the $[111]$ axes. The full dimer phase diagram is shown in Fig. \ref{R-phase} bottom.

For the monopole crystal such emergent dimer moves are generated by small transverse spin coupling compared with the nearest neighbour exchange. Application of degenerate perturbation theory \cite{Bergman2006}, yields a parameter ratio for the effective dimers of $\mu / g=0$, which is deep in the classical $R$-phase. The critical threshold  \cite{Moessner2001,Bergman2006} for entry into the quantum dimer liquid is
estimated numerically to be $\mu/g \sim 0.7$ \cite{Sikora2011}. As a consequence, we do not anticipate the appearance of an effective quantum dimer liquid in this system.
In addition, as the inclusion of dipolar corrections to the classical monopole model sees the system order into the spin-$R$ phase \cite{Jaubert2015} this would take a  putative quantum system, inclusive of dipolar interactions, even further from an emergent dimer liquid phase.

However, if one could push the system  into the quantum dimer liquid phase \cite{Pace2023}, the neutron scattering signature for the emergent field would strongly resemble that of quantum spin ice \cite{Benton2012}. The emergent field for the quantum dimers maps to lattice quantum electrodynamics (LQED) \cite{Sikora2011} with essentially the same structure and the same emergent photons which should show up in the inelastic neutron scattering spectrum.  Integrating over the photon bands to give static spin correlations, the pinch point structure of the classical system \cite{Brooks2014} would evolve. The dipolar correlations of the classical system map to correlations in four dimensional space time with projection onto three dimensions leading to a suppression of the pinch point intensities at the Brillouin zone centres. These predictions could be tested using configurations from the quantum Monte Carlo simulations of  Ref. \onlinecite{Sikora2011} and working backwards to construct the emergent transverse fragment of a monopole crystal. In this partial quantum liquid phase these modified spin correlations from the transverse fragment would coexist with the $[220]$ peaks from the longitudinal fragment or charge order. The intensity of these Bragg peaks should remain unchanged within the regime of emergent quantum dimer fluctuations.

\section{Discussion}

Any  vector field  can be separated, via a Helmholtz decomposition into divergence full (longitudinal), divergence free (transverse) and harmonic parts. In the monopole picture of spin ice and related materials, the magnetic moments play the role of an emergent lattice field in which such a decomposition or magnetic moment fragmentation is of particular interest. In this description, both ground state and excitation spectrum separate perfectly into elements from the different fragments with well separated energy scales. The monopoles \cite{Castelnovo2008,Ryzhkin2005} are built from the longitudinal fragment and are high energy objects, the classical macroscopic degeneracy or quantum photon spectrum come from the transverse fragment, while the topological properties \cite{Moessner2003,Jaubert2010,Turrini2022,Pili2022} are controlled by the harmonic fragment.

We have shown here that it is extremely useful to carry this decomposition through to the analysis of neutron scattering results, as each component gives a distinct contribution to the neutron scattering intensity.  Previous texts have concentrated on situations in which an ordered monopole fragment coexists with the transverse fragment in the form of a correlated spin liquid \cite{Brooks2014,Lefrancois2017,Canals16,Cathelin2020,Paddison2016}. Here we show that such systems with magnetic charge ordering, when driven into a fully ordered phase, either through quantum fluctuations or by small corrections to the monopole picture, form fragmented double-$q$ structures in which each fragment orders with a distinct ordering wave vector. Due to the separation in energy scales, quantum fluctuations are largely restricted to the transverse fragment. In consequence  the intensity reduction of the transverse fragment compared to a known classical limit can be used as a diagnostic tool for the level of quantum fluctuations.

In the specific examples chosen; the KII phase of kagom\'e ice and the monopole crystal of spin ice, the transverse fragment maps onto a $Z_2$ sector of the emergent field for a hard core dimer system on hexagonal and diamond lattices respectively,  so that the neutrons indirectly probe dimer solids, both classical and quantum.
The analysis we propose relies on the existence of a gapped energy spectrum above the ground state. In this case, the proposed quantum resonances of spins, or effective dimers around small closed loops will lead to a quantifiable reduction in the observed neutron scattering intensity. For this to hold, both the temperature scale and the neutron energy resolution must be smaller than this gap.

Bojesen and Onoda  \cite{Bojesen2017} have argued that their quantum Monte Carlo data for spin ice in a modest $[111]$ field are consistent with the development of an emergent quantum dimer solid at low temperature. Our paper provides a protocol for a detailed analysis allowing for the distinction between the quantum phase and its classical analogue. The energy scale associated with this quantum phase is extremely low; between $1/20$  and $1/320$ of the nearest neighbour coupling strength, so that quantitative measurement appears to be at the limit of numerical resolution. However, a clearer quantum limit is reached in dedicated quantum dimer simulations on a hexagonal lattice \cite{Schlittler2017}. Our protocol could be tested in detail from such simulations by reconstructing a single $Z_2$ sector of the emergent field from the dimers and constructing the corresponding neutron scattering plots.

The low energy scales associated with quantum spin ice have so far made identification of experimental systems extremely difficult. One promising example is Pr$_2$Hf$_2$O$_7$ \cite{Sibille2018} which shows some evidence of a quantum spin liquid ground state from inelastic neutron scattering of single crystal samples. Precision experiments in a $[111]$ field would certainly be of interest here as the first stage in the quest to observe dimensional reduction to the two-dimensional quantum phases predicted in Ref. \onlinecite{Bojesen2017} and discussed in detail above. The stacked kagom\'e layer material Ho$_3$Mg$_2$Sb$_3$O$_{14}$ appears to show quantum corrections to a classical fragmented magnetic structure closely related to the KII phase of kagom\'e ice \cite{Dun2020}, although for the moment only powder samples exist and the synthesis of pure samples appears challenging. In the absence of single crystals, our analysis could be extended to treat a powder sample. This would be of interest as the signal from the quantum spin$-P$ phase introduced above would be distinct from the alternative quantum phases predicted by Dun {\it{et al.}} \cite{Dun2020}.
However, at least in the short term, artificial systems, such as cold atom ice fabricated from Rydberg atoms \cite{Glaetzle2014} could hold the advantage over materials and could provide promising options for the observation of tuneable quantum fluctuations in systems with ice geometry.

Looking forward, open questions remain for the thermal to quantum crossover for the phase transitions from ordered to spin liquid phases. In two dimensions in the $g=0$ limit of eqn. (\ref{hex-flip}), the thermal phase transition from the $\sqrt{3}\times\sqrt{3}$ to KII phase should map to a roughening transition and hence  be of Kosterlitz-Thouless type \cite{Alet2006}, although this could change in the presence of monopole defects \cite{Chern2011,Wang2020}. As quantum fluctuations are switched on the fate of the topological transition is far from clear and open to further studies. In three dimensions an evolution of tri-critical form is predicted, taking the thermal $R$-phase to dimer liquid transition from topological to first order as quantum fluctuations increase \cite{Bergman2006}.
In the context of this paper, the ultimate goal would be to prepare experimental signatures of these subtle questions through use of the fragmentation picture in the neutron scattering analysis.

As a final note, the more mathematically inclined reader will notice that the Helmholtz decomposition can be expanded further with regard to the transverse term. Any divergence-free vector field can be decomposed into toroidal and poloidal fields:
\begin{align}
    \curl{\vec {A}} & = \vec {T} + \vec {P}\nonumber                                   \\
                    & = \grad{\phi} \cross \hat{r} + \curl(\grad{\chi} \cross \hat{r})
\end{align}
where $\hat{r}$ is a radial unit vector, $\phi$ is the toroidal and $\chi$ the poloidal scalar potentials.
Together with the longitudinal potential $\Psi$ they make up the Debye potentials and allow the mapping of any vector field onto a set of three scaler fields, up to a harmonic contribution \cite{DubovikTugushev1990,Spaldin2008}.
The fields from a single point dipole are purely poloidal while toroidal fields are characteristic of circular solenoids or toroids. The complete  decomposition of the transverse magnetic fragment into poloidal and toroidal elements is beyond the scope of this discussion but pragmatically one can assume that the extensive loop network leading to pinch point scattering patterns is due largely to the poloidal component, while short loops contain a toroidal contribution. In modified spin ice models with induced attractive interactions between monopoles of like charge, low energy excitations include like charge clusters characterised by loops of spin flips, identified as toroidal loops \cite{UdagwaJaubert2016,Tomonari2018,Kiese2022-PP}. Using the fragmentation picture it is straightforward to show that such clusters lead to isolated loops in the transverse fragment, which indeed correspond to a pure toroidal contribution. In a spin liquid phase dominated by such loops, the diffuse neutron scattering is characterised by half-moons of high intensity straddling the Brillouin zone centre, rather than the pinch points of the Coulomb phase. This strongly suggests that magnetic moment fragmentation could be an essential tool for a complete description of such systems.

\begin{acknowledgments}
    We acknowledge financial support from the ``Agence Nationale de la Recherche'' under Grant No. ANR-19-CE30-0040. We thank Shigeki Onoda for authorising the reproduction of our Figure (\ref{Onoda}) from reference [\onlinecite{Bojesen2017}], as well as Baptiste Bermond for fruitful discussions.
\end{acknowledgments}

\bibliography{Fragment-neutrons}

\end{document}